# Stochastic Tissue Window Normalization of Deep Learning on CT

Yuankai Huo [1*], Yucheng Tang [1], Yunqiang Chen [2], Dashan Gao [2], Shizhong Han [2], Shunxing Bao [1], Smita De [3], James G. Terry [4], J. Jeffery Carr [4], Richard G. Abramson [5], and Bennett A. Landman [1,4]

[1] Vanderbilt University, Department of Electrical Engineering and Computer Science,
[2] 12 Sigma Technologies
[3] Cleveland Clinic
[4] Vanderbilt University Medical Center, Department of Radiology

1 *Abstract*—Tissue window filtering has been widely used in deep learning for computed tomography (CT) image analyses to improve training performance (e.g., soft tissue windows for abdominal CT). However, the effectiveness of tissue window normalization is questionable since the generalizability of the trained model might be further harmed, especially when such models are applied to new cohorts with different CT reconstruction kernels, contrast mechanisms, dynamic variations in the acquisition, and physiological changes. In this paper, we evaluate the effectiveness of both with and without using soft tissue window normalization on multi-site CT cohorts. Moreover, we propose a new stochastic tissue window normalization (SWN) method to improve the generalizability of tissue window normalization. Different from the naïve random sampling, the SWN method centers the randomization around the soft tissue window to maintain the specificity for abdominal organs. To evaluate the performance of different strategies, 80 training and 453 validation and testing scans from six datasets are employed to perform multi-organ segmentation using standard 2D U-Net. The six datasets cover the scenarios where the training and testing scans are from (1) same scanner and same population, (2) same CT contrast but different pathology, and (3) different CT contrast and pathology. The traditional soft tissue window and non-windowed approaches achieved better performance on (1). The proposed SWN achieved general superior performance on (2) and (3) with statistical analyses, which offers better generalizability for a trained model.

*Index Terms* — Tissue Window, CT, Deep Learning, Segmentation

## I. Introduction

Computed tomography (CT) is a quantitative imaging technique that produces imaging intensities normalized in Hounsfield Units (HU) (e.g., air as -1000 HU, water as 0 HU). The quantitative meaning of intensity units allows clinical

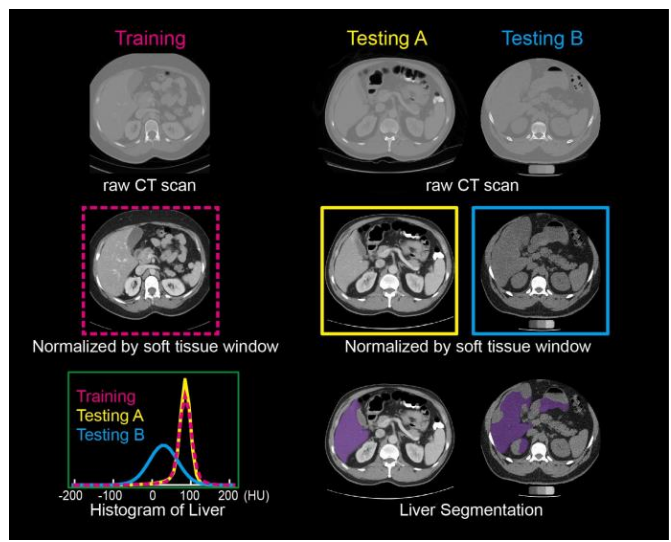

Fig. 1. The soft tissue window normalization works well when the distribution of the testing scan (Testing A) matches the training scan. However, the performance might be degraded on the testing scan (Testing B) with different CT contrast. The mechanism of modifying the contrast is to apply a soft tissue window (-160<HU<240) on the raw CT scans.

practitioners to define typical window ranges (e.g., the range of intensities to display) to enhance the visual contrasts for particular tissues or organs by applying tissue windows [1]. A tissue window is an intensity band-pass filter, which only keeps the intensities within the band and censors the intensities beyond the maximal/minimal values. The band is commonly decided according to the HU of targeting organ. For instance, a lung window (-1150<HU<350) is typically applied to investigate lung images [2], and a soft tissue window (-160<HU<240) is commonly employed to enhance the image contrast for abdominal organs [1]. Tissue windows not only improve the image contrast for human visualization [3] but also

The authors of the paper are directly employed by the institutes or companies provided in this paper. This research was supported by NSF CAREER 1452485, NIH grants 5R21EY024036, R01EB017230, 1R21NS064534, 1R03EB012461, R01 DK113980, 6R01 DK112262. Yuankai Huo, Yucheng Tang, Richard G. Abramson, and Bennett A. Landman are supported by the Vanderbilt-12 Sigma Research Grant (Huo/Abramson/Landman). Richard G. Abramson is also receiving partial support from 2U01CA142565 and P30 CA068485. J. Jeffery Carr and James G. Terry are in part supported by R01 DK113980, DK Locke (PI) (09/01/17-08/31/22) "CKD risk prediction among obese living kidney donors". This project will evaluate novel biomarkers of risk as relates to obese living kidney donors. J. Jeffery Carr and James G. Terry are co-Investigator 6R01 DK112262, NIDDK Koethe (PI) (02/01/17-01/31/22) "The role of adipose-resident T cells in HIV-associated glucose intolerance". No conflicts of interest, financial or otherwise, are declared by Yunqiang Chen, Dashan Gao, Shizhong Han, Shunxing Bao, and Smita De. We thank Naiyun Zhou for helping organize part of the data.

Corresponding Author: Yuankai Huo, E-mail: yuankai.huo@vanderbilt.edu

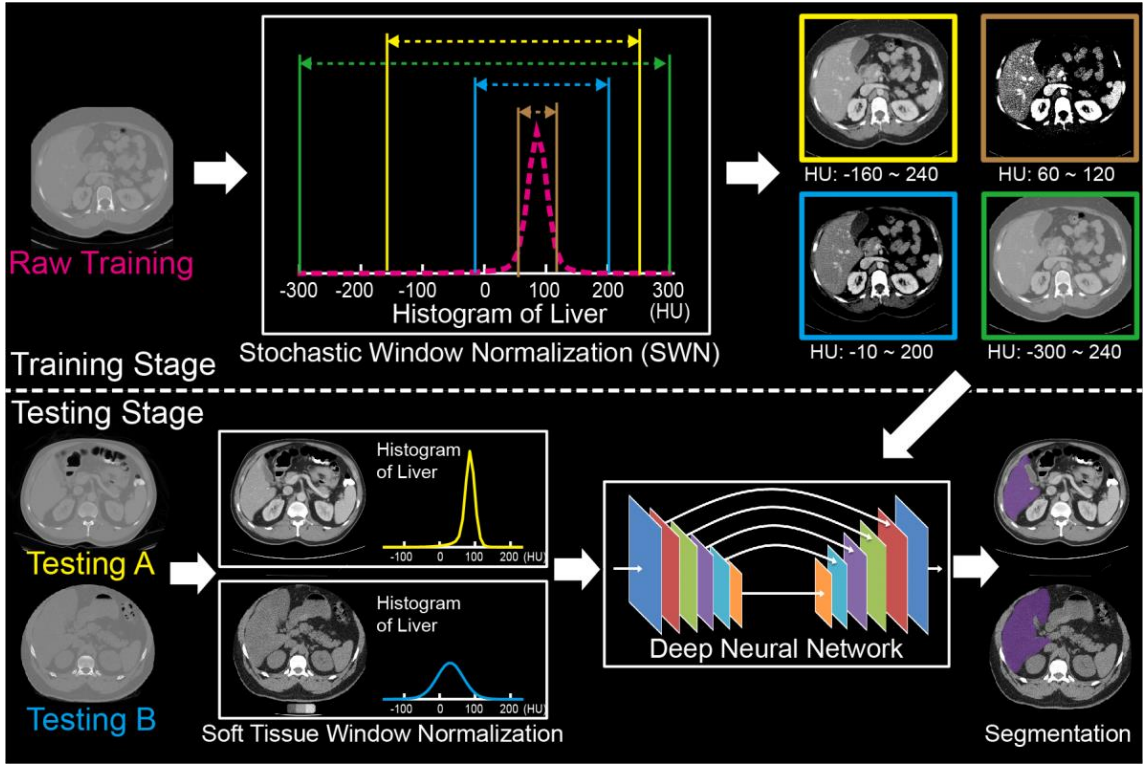

Fig. 2. The workflow of deploying the proposed stochastic tissue window normalization (SWN) to train a standard 2D U-Net segmentation network.

filter out texture/noise in unrelated tissues, organs, and background.

In recent years, the tissue window filtering process has been widely adapted to deep learning methods on CT image analyses [4-7]. The rationale of using tissue window normalization (preprocessing) is to get rid of the unnecessary information before the machine learning stage, which enhances the specificity of the trained deep learning model. The "specificity" in this study is referred to the performance of deploying a trained deep learning network on testing data with the same imaging acquisition as the training data. The hypothesis behind that is the HU values are standardized and homogenous across different cohorts. However, this hypothesis might not always be valid for some imaging scenarios, including but not limited to, (1) different CT hardware, (2) potential confounds of CT reconstruction kernels, (3) different contrast-enhanced CT imaging, (4) dynamic variations in acquisition, (5) physiological changes, et cetera. As a result, the generalizability of the trained model using fixed tissue window might be degraded when it is applied to the heterogeneous clinical CT scans (Figure 1). The "generalizability" in this study is defined as the performance of deploying a trained deep learning network on testing data with the different imaging acquisition from the training data.

In this paper, we investigate the effectiveness of standard soft tissue window normalization (STN) for canonical multi-organ segmentation task compared with whole intensity range (WIR, without using tissue windows). Moreover, we propose a new stochastic tissue window normalization (SWN) method to leverage the generalizability upon STN. Different from naively using random windows, we limit the window variations to be centralized around the soft tissue window to improve specificity.

Eighty non-contrast CT scans with healthy organs are used to train a standard 2D U-Net from [8]. Then, 20 scans from the same cohort and 433 scans from different cohorts are used to evaluate the effectiveness of STN, WIR and SWN, which covers the scenarios where the training and testing scans are from (1) same scanner and same population, (2) same CT contrast but different pathology, and (3) different CT contrast and pathology.

## II. METHOD

### A. Stochastic Tissue Window Normalization

Figure 2 demonstrates the principle of training an organ segmentation network using SWN, which randomly samples the window size and location beyond the STN. A tissue window is determined by two parameters: window level (center) and window size [1]. Instead of only pursuing generalizability by natively sampling random windows, we force the randomly sampled windows to be centered around soft tissue window to maintain the specificity. To achieve that, we used the soft tissue window (window level $L = 40$, half window size $W = 200$) as the centers of the random sampling. The pseudo code of the proposed SWN method is provided in Figure 3. Briefly, we employed two Gaussian distributions to add variability upon the soft tissue window. The new windows are randomly sampled from the following two Gaussian distributions,

```
Pseudo Code of the Stochastic Tissue Window Normalization (SWN)
────────────────────────────────────────────────────────────────
SWN(image, x, y):
    image = float(image)                        // convert image volume to float
    L = gaussian_random(μ = 40, σ = x)          // generate random level
    W = gaussian_random(μ = 200, σ = y)         // generate random window
    W = abs(W)                                  // avoid the negative window
    max_threshold = L + W                       // maximum threshold of window
    min_threshold = L − W                       // minimum threshold of window
    //windowing
    image[image > max_threshold] = max_threshold
    image[image < min_threshold] = min_threshold
    image = 255 * (image − min_threshold) / (max_threshold − min_threshold)
```

Fig. 3. Pseudo-Code of the Stochastic Tissue Window Normalization (SWN). The terms are defined based on Eq (1), (2), and (3).

$$L \sim Gaussian\ (\mu = 40, \sigma = x) \quad (1)$$
$$W \sim Gaussian\ (\mu = 200, \sigma = y) \quad (2)$$

where $x$ and $y$ are the two coefficients to control the variabilities of the random windows. In the paper, we used the format "$[x, y]$" to show the values of $x$ and $y$ for any experiments performed by SWN. During the training, a 2D training image slice $I_i$ is normalized by the sampled window with the following steps:

$$\begin{aligned} I_i(I_i > (L_i + W_i)) &= (L_i + W_i) \\ I_i(I_i < (L_i - W_i)) &= (L_i - W_i) \\ I'_i &= \frac{I_i - (L_i - W_i)}{2W_i} \end{aligned} \quad (3)$$

Note that, the $L_i$ and $W_i$ are different for each input during training, which are randomly sampled from the aforementioned two Gaussian distributions. For WIR, the intensity with in whole major intensity range (-1000<HU<1000) are normalized for training without applying any tissue windows. In the testing stage, we preprocess every testing scan using standard soft tissue window for STN and SWN, while not using such window for WIR.

B. *Multi-organ Segmentation Network*

To evaluate the effectiveness of using tissue window normalization, we keep the training network and processing standardized. The canonical 2D U-Net [8] is employed as the base network. The same data augmentation stages (random cropping, padding, rotation, translation) are performed to enhance the spatial generalizability. First, all input CT image voxels are converted to floating point numbers with 32 bits ("float"). Then all the input 2D CT images (after windowing and preprocessing) are further normalized to 0 to 255 ("float") with resolution $512 \times 512$. The number of input channels is one, while the number of output channels is eight (including background, spleen, right kidney, left kidney, liver, stomach, pancreas, body mask). The Adam optimizer [9] with learning rate 0.00001 is used with a batch size of six. Weighted cross-entropy is used as the loss function, whose weights of eight channels are [1, 10, 10, 10, 5, 10, 10, 1]. The models are trained with the maximum of 100 epochs. When training each epoch, every image is windowed once, across different windowing methods. The level and window size are randomly decided for each time when using the proposed SWN. Therefore, the windows are different, even for the same image across different epochs. During testing stage, the standard soft-tissue window (without randomness) is used for SWN to have a fair comparison with the STW method. The learning rate, epoch number, and the weights were optimized from internal validation and were applied to all testing cohort consistently. Notably, the same hyper-parameters are used across all experiments, except the tissue window normalization.

C. *Training and Validation Data (Same Scanner and Population)*

MLBCV (Multi-organ): 100 abdominal CT scans were obtained from the MICCAI 2015 Multi-atlas Labeling Beyond the Cranial Vault (MLBCV) challenge [10]. The data were acquired from portal venous phase CT modality with variable volume sizes ($512 \times 512 \times 33$ to $512 \times 512 \times 158$) and field of views (approx. $300 \times 300 \times 250$ mm3 to $500 \times 500 \times 700$ mm3). The in-plane resolution varies from $0.54 \times 0.54$ mm2 to $0.98 \times 0.98$ mm2. Among 100 scans, 80 were used as training while the remaining 20 were used as validation. Six organs (spleen, right kidney, left kidney, liver, stomach, pancreas) from MLBCV are used as training targets.

| | Training | Validation | Testing (Same CT Contrast as Training) | | | Testing (Different Contrast) | |
|---|---|---|---|---|---|---|---|
| Cohort | MLBCV | MLBCV | Decathlon | LiTS | FNH | AADHS | DelayedCT |
| Modality | portal venous phase CT | portal venous phase CT | portal venous phase CT | portal venous phase CT | portal venous phase CT | Non-contrast | Delayed CT |
| Organ | Multi-organ | Multi-organ | Pancreas | Liver | Liver | Liver | Kidneys |
| Pathology | Healthy | Healthy | with Tumor | with Tumor | FNH | Fatty Liver | Heathy |
| Scans # | 80 scans | 20 scans | 282 scans | 131 scans | 8 scans | 5 scans | 5 scans |

Fig. 4. Summary of training, validation, and testing cohorts

## III. DATA

### A. Testing Data (Same CT Contrast but Different Pathology)

Figure 4 summarizes the six datasets used in this study. All datasets were acquired in deidentified form under institutional review board approval.

**Decathlon (Pancreas):** 282 abdominal CT scans with manual pancreas segmentation were obtained from MICCAI 2018 Medical Segmentation Decathlon (Pancreas Tumor) dataset. The data were acquired from portal venous phase CT modality. The details of the data can be found at http://medicaldecathlon.com.

**LiTS (Liver):** 131 abdominal CT scans with liver manual segmentation were obtained from Liver Tumor Segmentation (LiTS) Challenge. The data were acquired from portal venous phase CT modality. The details of the data can be found at https://competitions.codalab.org/competitions/17094.

**FNH (Liver):** 8 abdominal CT scans with liver manual segmentation were internally acquired from patients with Focal Nodular Hyperplasia (FNH) lesion. The data were acquired from contrast-enhanced in portal venous phase CT modality with in-plane image size $512 \times 512$ and resolution from 0.5 mm to 0.8 mm. The slice thickness is 5 mm.

### B. Testing Data (Different CT Contrast and Pathology)

**AADHS (Liver):** 5 abdominal CT scans with fatty liver diagnosis and manual liver segmentations were obtained from African American-Diabetes Heart Study (AADHS) dataset. The data were acquired from non-contrast CT modality with in-plane resolution $512 \times 512$. The details of the data can be found at [11].

**Delayed (Kidneys):** 5 abdominal CT scans with manual left and right kidney segmentation were acquired internally with excretory phase sequences. The scans were performed in the prone position at an 8 min delay per institutional protocol with 3 mm axial reconstructions.

## IV. SIMULATION

**Specificity and Generalizability Analysis.** The 20 validation CT scans were used to evaluate the specificity and generalizability of STN, WIR, and SWN. To test the specificity and generalizability, we performed a simulation, which adds or subtract constant values on 20 validation scans (from -300 to +300 in steps of 25 HU). That experiment simulates the intensity variations in testing data when applying the trained model. The 20 validation CT scans were used since the data were acquired from the same scanner as the training data. Therefore, the spatial effects will be minimized and the difference in performance is solely from the global variations on intensities. Figure 5 shows the variations of segmentation performance on six organs with the changes in raw intensities.

## V. EMPIRICAL VALIDATION

The 20 MLBCV scans are used to evaluate the performance of different window normalization strategies for the scenarios that the training and testing scans are from the "same scanner and population". The Dice similarity coefficient (DSC) has been used as the metrics to show the segmentation accuracy.

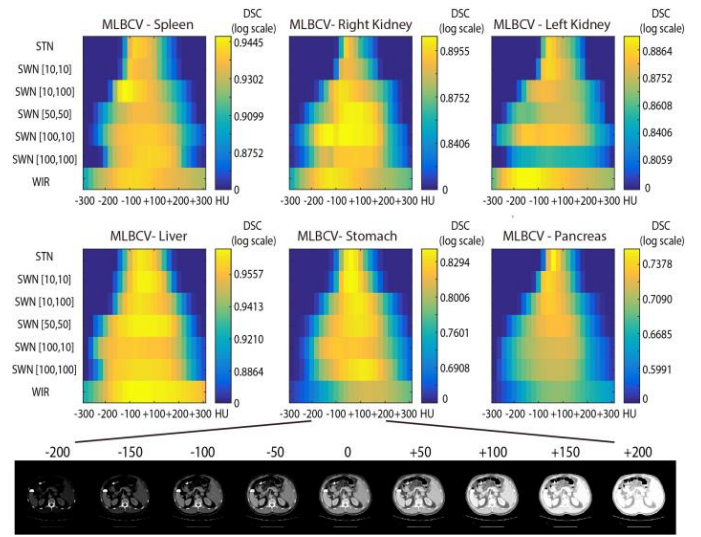

Fig. 5. This figure shows the specificity and generalizability of STN, WIR, and SWN. To test the different tissue window normalization strategy, the testing scans have been added or subtract-ed constant values and fed into the same network. The color indicates the mean Dice values across 20 validation scans for each organ. The width of the yellow color range in each row shows the generalizability, while the brightness indicates the specificity. The proposed SWN has better generalizability compared with STN, and better specificity compared with WIR.

The Decathlon, LiTS, and FNH cohorts are employed to evaluate the performance of different window normalization strategies for the scenarios that the training and testing scans are from "same CT contrast but different pathology".

The AADHS and Delayed cohorts are employed to evaluate the performance of different window normalization strategies for the scenarios that the training and testing scans are from "different CT contrast and pathology".

### A. Internal Validation (MLBVC)

The qualitative and quantitative results of 20 MLBVC validation scans are shown in Figure 6 and 7 respectively. The detailed measurements of six labels are presented in Table 1. As the training and validation datasets are from the same cohort and the same scanner, the intensities of training scans and testing scans are homogeneous. Therefore, the canonical STN or WIR methods achieved superior performance in either median DSC or mean DSC for all six organs. In Table 1, The best DSC results are marked as bold. Briefly, the greater median and mean DSC indicate the better segmentation performance referring to the manual segmentations. The smaller standard deviation (STD) of DSC means the variation of the segmentation performance is smaller and more consistent across the cases. The symbol "$-$" indicates that the difference between the corresponding method and the reference method ("Ref.") is not significant. The symbol "↑" and "↓" means significantly higher and lower respectively using the Wilcoxon signed rank test with $p<0.05$. The symbol "*" means the false discovery rate (FDR) corrected p value within the corresponding abdominal organ is $< 0.05$, with number of

comparisons = 12 of each organ.

### B. *5.2 External Validation on Same Imaging Protocol*

We group the results of Decathlon, LiTS, and FNH as the external validation results on same CT modality, since such datasets were acquired from the same imaging protocol (portal venous phase) as the training datasets but from different sites. The qualitative and quantitative results of different methods are presented in Figure 8 and 9. The corresponding detailed measurements are provided in Table 2. When performing the trained model on external validation datasets with the same imaging protocol but different sites and pathologies, the proposed SWN method achieved superior performance compared with the canonical STN and WIR methods.

### C. *5.3 External Validation on Different Imaging Protocol*

The trained model from portal venous phase CT scans is evaluated using the non-contrast CT scans (AADHS) and delayed phase CT scans (Delayed). In this scenario, the HU intensities of livers in AADHS are systematically different from training data. Meanwhile, the HU intensities of kidneys in Delayed are systematically different from training data. Therefore, the intensities of targeting organs in training and testing datasets are heterogeneous. The qualitative and quantitative results are presented in Figure 10 and 11. The corresponding detailed measurements are provided in Table 3. From the results, the proposed SWN method achieved superior performance compared with the canonical STN and WIR methods.

## VI. CONCLUSION AND DISCUSSION

We evaluate the effectiveness of both tissue window normalization and non-windowed methods for deep learning on CT organ segmentation tasks. The soft tissue window typically yields superior performance on segmenting smaller and more challenging organs (pancreas and stomach). Meanwhile. the segmentation performance of without using tissue window techniques achieved superior performance on larger and easier organs (liver and spleen).

From internal validation (training and testing data are from the same scanner and population), the STN and WIR achieved overall better segmentation performance (Figure 7 and Table 1). We propose a new stochastic tissue window normalization method and evaluate the STN, WIR and SWN methods using simulation (Figure 5) different external testing cohorts.

According to the absolute differences in Dice values (highlighted in Bold), the propose SWN method achieved generally better Dice scores, when evaluated on the testing scans acquired from the different scanner but same contrast (Figure 9 and Table 2), When evaluated on the testing scans acquired from different modalities and different pathologies (Figure 11 and Table 3), the proposed SWN method also achieved generally superior Dice values compared with STN and WIR. The proposed SWN provided better generalizability of a trained model while preserving the specificity compared with STN and WIR.

The standard Wilcoxon signed-rank test statistical analyses (highlighted with colors) is used in the study. When the training and testing scans are regimented to be acquired from the same scanner, protocol, and patient population (Table 1), the proposed method demonstrates improved benchmarks as compared to the standard method. It means the simple standard intensity normalization methods are more proper for the internal validation. But in the real world, we typically would like to train a more generalizable deep learning model, which can be applied directly to different cohorts and populations (Table 2 and 3). Under such external validation scenarios, the generalizability of the trained model is essential, especially when the number of available training cases are typically in small-scale for medical imaging applications. The proposed method achieves overall superior performance when the testing and training cohorts are more heterogeneous, which leverages the segmentation performance of the trained models on the different testing imaging protocols. Under the more restricted scenarios, FDR correction is applied to correct the original p-values for multiple comparison (highlighted with "*"). After FDR correction, the differences for MLBCV-spleen, MLBCV-stomach (Table 1), FNH-liver (Table2), AADHS-liver, Delayed-left kidney and Delayed-right kidney (Table 3) are not significant. The non-significant comparisons in Table 2 and 3 are due to the relatively small sizes of available cohorts (i.e., 5 to 8 patients).

The standard 2D U-Net is employed as the segmentation network to evaluate the performance of using tissue windows. While this combination is successful, we do not claim optimality of using 2D U-Net. To achieve the superior segmentation network is not the major aim of this work. In the future, it would be also interesting to have the organs from different contrasts labeled by different human expert. In that case, the inter-rater reliability is able to be calculated, which can be used to evaluate the automatic detection with human variability.

The proposed method is validated on the soft tissue window. However, other types of tissue windows (e.g., lung, cardiac, liver window etc.) have also been widely used in different applications. Theoretically, the stochastic tissue window would also improve the generalizability of deep network for such applications. Therefore, it would be useful to extend and validate the proposed method to such applications in the future. Another limitation of the proposed window based normalization is that it sacrifices the physical information behind the HU standardization.

**SPIE Author Biography**
First Author is a research assistant professor at the Vanderbilt University. He received his BS degree in Telecommunication Engineering from Nanjing University of Posts and Telecommunications in 2008, his MS degrees in Information and Telecommunication Engineering and Computer Science from the Southeast University and the Columbia University in 2011 and 2014 respectively, and his PhD degree in Electrical Engineering from the Vanderbilt University in 2018. He is the author of more than 50 journal and conference papers in medical image analysis. He is a member of SPIE.

**Table 1.** Segmentation performance on MLBCV

| | STN | WIR | SWN [10,10] | SWN [10,100] | SWN [50,50] | SWN [100,10] | SWN [100,100] |
|---|---|---|---|---|---|---|---|
| **MLBCV - Spleen** | | | | | | | |
| Median | 0.9438 | **0.9469** | 0.9444 | 0.9456 | 0.9437 | 0.9442 | **0.9469** |
| Mean | 0.9380 | 0.9407 | 0.9359 | 0.9391 | 0.9368 | 0.9399 | **0.9413** |
| Std | 0.0317 | 0.0294 | 0.0333 | 0.0247 | 0.0283 | **0.0194** | 0.0279 |
| p<0.05 | Ref. | — | p=0.048 ↓ | — | — | — | — |
| p<0.05 | — | Ref. | — | — | — | — | — |
| **MLBCV - Right Kidney** | | | | | | | |
| Median | **0.9307** | 0.9274 | 0.9194 | 0.9252 | 0.9240 | 0.9288 | 0.9187 |
| Mean | 0.8898 | 0.8942 | 0.8894 | 0.8948 | 0.8995 | **0.8999** | 0.8936 |
| Std | 0.1231 | 0.1046 | 0.0966 | 0.0894 | **0.0801** | 0.0887 | 0.0859 |
| p<0.05 | Ref. | — | * p=0.004 ↓ | — | — | — | — |
| p<0.05 | — | Ref. | * p=0.006 ↓ | — | — | — | — |
| **MLBCV - Left Kidney** | | | | | | | |
| Median | 0.9360 | **0.9400** | 0.9364 | 0.9337 | 0.9251 | 0.9346 | 0.9132 |
| Mean | 0.8840 | **0.8859** | 0.8855 | 0.8801 | 0.8762 | 0.8835 | 0.8571 |
| Std | 0.2087 | 0.2096 | 0.2093 | 0.2083 | 0.2074 | 0.2089 | **0.2040** |
| p<0.05 | Ref. | — | — | — | * p=0.003 ↓ | — | * P<0.001 ↓ |
| p<0.05 | — | Ref. | — | — | * p=0.003 ↓ | — | * P<0.001 ↓ |
| **MLBCV – Liver** | | | | | | | |
| Median | 0.9633 | 0.9659 | **0.9662** | 0.9633 | 0.9648 | 0.9577 | 0.9634 |
| Mean | 0.9622 | **0.9646** | 0.9639 | 0.9613 | 0.9640 | 0.9575 | 0.9611 |
| Std | 0.0096 | **0.0086** | 0.0089 | 0.0110 | **0.0086** | 0.0127 | 0.0103 |
| p<0.05 | Ref. | * p=0.010 ↑ | * p=0.025 ↑ | — | — | * P<0.001 ↓ | — |
| p<0.05 | * p=0.010 ↓ | Ref. | — | * p=0.011 ↓ | — | * P<0.001 ↓ | * p=0.010 ↓ |
| **MLBCV - Stomach** | | | | | | | |
| Median | **0.8528** | 0.8102 | 0.8412 | 0.8418 | 0.8348 | 0.8306 | 0.8325 |
| Mean | 0.8377 | 0.8029 | **0.8380** | 0.8327 | 0.8305 | 0.8234 | 0.8307 |
| Std | 0.0805 | 0.1052 | **0.0777** | 0.0995 | 0.0891 | 0.0944 | 0.0845 |
| p<0.05 | Ref. | p=0.019 ↓ | — | — | — | — | — |
| p<0.05 | p=0.019 ↑ | Ref. | p=0.014 ↑ | p=0.019 ↑ | — | — | — |
| **MLBCV - Pancreas** | | | | | | | |
| Median | **0.7620** | 0.7196 | 0.7453 | 0.7407 | 0.7234 | 0.7294 | 0.7336 |
| Mean | **0.7483** | 0.7030 | 0.7357 | 0.7344 | 0.7313 | 0.7215 | 0.7167 |
| Std | 0.1149 | 0.1140 | **0.1038** | 0.0886 | 0.1091 | 0.1279 | 0.1046 |
| p<0.05 | Ref. | * p=0.007 ↓ | — | * p=0.007 ↓ | — | * p=0.003 ↓ | * p=0.005 ↓ |
| p<0.05 | * p=0.007 ↑ | Ref. | * p=0.012 ↑ | * p=0.017 ↑ | * p=0.033 ↑ | — | — |

The best DSC results are marked as bold. The symbol "—" indicates that the difference between the corresponding method and the reference method ("Ref.") is not significant. The symbol "↑" and "↓" means significantly higher and lower respectively using the Wilcoxon signed rank test with p<0.05. "*" means the FDR corrected p value is also < 0.05.

**Table 3.** Performance on Testing Data (Same CT Contrast, Different Pathology)

|  | STN | WIR | SWN [10,10] | SWN [10,100] | SWN [50,50] | SWN [100,10] | SWN [100,100] |
|---|---|---|---|---|---|---|---|
| | | | **Decathlon - Pancreas** | | | | |
| Median | 0.6908 | 0.6407 | **0.6996** | 0.6880 | 0.6933 | 0.6870 | 0.6972 |
| Mean | 0.6480 | 0.6009 | **0.6714** | 0.6612 | 0.6607 | 0.6590 | 0.6665 |
| Std | 0.1639 | 0.1645 | 0.1432 | **0.1403** | 0.1507 | 0.1467 | 0.1416 |
| p<0.05 | Ref. | * p<0.001 ↓ | * p<0.001 ↑ | — | * p=0.034 ↑ | — | * p=0.011 ↑ |
| p<0.05 | * p<0.001 ↑ | Ref. | * p<0.001 ↑ | * p<0.001 ↑ | * p<0.001 ↑ | * p<0.001 ↑ | * p<0.001 ↑ |
| | | | **LiTS - Liver** | | | | |
| Median | 0.9396 | 0.9425 | 0.9414 | 0.9420 | **0.9439** | 0.9389 | 0.9425 |
| Mean | 0.9321 | 0.9294 | 0.9315 | **0.9351** | 0.9335 | 0.9288 | 0.9346 |
| Std | 0.0307 | 0.0472 | 0.0405 | **0.0300** | 0.0376 | 0.0398 | **0.0300** |
| p<0.05 | Ref. | — | — | * p=0.015 ↑ | * p=0.009 ↑ | * p=0.008 ↑ | * p=0.015 ↑ |
| p<0.05 | — | Ref. | — | — | — | * p=0.011 ↓ | — |
| | | | **FNH - Liver** | | | | |
| Median | 0.9317 | 0.9395 | 0.9389 | 0.9430 | 0.9422 | 0.9408 | **0.9443** |
| Mean | 0.9295 | 0.9367 | 0.9386 | 0.9408 | **0.9423** | 0.9361 | 0.9399 |
| Std | 0.0264 | 0.0181 | 0.0138 | **0.0119** | 0.0139 | 0.0203 | 0.0166 |
| p<0.05 | Ref. | — | — | — | p=0.008 ↑ | — | p=0.016 ↑ |
| p<0.05 | — | Ref. | — | — | — | — | — |

The best DSC results are marked as bold. The symbol "—" indicates that the difference between the corresponding method and the reference method ("Ref.") is not significant. The symbol "↑" and "↓" means significantly higher and lower respectively using the Wilcoxon signed rank test with p<0.05. "*" means the FDR corrected p value is also < 0.05.

**Table 2.** Performance on Testing Data (Different CT Contrast and Pathology)

|  | STN | WIR | SWN [10,10] | SWN [10,100] | SWN [50,50] | SWN [100,10] | SWN [100,100] |
|---|---|---|---|---|---|---|---|
| | | | **AADHS - Liver** | | | | |
| Median | 0.8290 | 0.8752 | 0.8983 | **0.9017** | 0.8924 | 0.8433 | 0.8892 |
| Mean | 0.7799 | 0.8214 | 0.8458 | **0.8811** | 0.8797 | 0.8084 | 0.8589 |
| Std | 0.1661 | 0.1248 | 0.1266 | **0.0559** | 0.0449 | 0.1433 | 0.1039 |
| p<0.05 | Ref. | — | p<0.05 ↑ | — | p<0.05 ↑ | p<0.05 ↑ | p<0.05 ↑ |
| p<0.05 | — | Ref. | p<0.05 ↑ | — | p<0.05 ↑ | — | p<0.05 ↑ |
| | | | **Delayed - Right Kidney** | | | | |
| Median | 0.8847 | 0.8875 | 0.8678 | 0.9048 | **0.9084** | 0.8954 | 0.8905 |
| Mean | 0.8652 | 0.8673 | 0.8690 | **0.9035** | 0.9031 | 0.8921 | 0.8995 |
| Std | 0.0352 | 0.0719 | 0.0526 | 0.0341 | 0.0256 | 0.0281 | **0.0245** |
| p<0.05 | Ref. | — | — | p<0.05 ↑ | p<0.05 ↑ | — | — |
| p<0.05 | — | Ref. | — | — | — | — | — |
| | | | **Delayed – Left Kidney** | | | | |
| Median | 0.8755 | 0.8328 | 0.8491 | **0.8994** | 0.8841 | 0.8853 | 0.8936 |
| Mean | 0.8580 | 0.7898 | 0.8359 | **0.8987** | 0.8910 | 0.8818 | 0.8913 |
| Std | 0.0605 | 0.1010 | 0.0427 | 0.0334 | 0.0298 | **0.0272** | 0.0293 |
| p<0.05 | Ref. | p<0.05 ↓ | — | — | — | — | — |
| p<0.05 | p<0.05 ↑ | Ref. | — | — | — | — | p<0.05 ↑ |

The best DSC results are marked as bold. The symbol "—" indicates that the difference between the corresponding method and the reference method ("Ref.") is not significant. The symbol "↑" and "↓" means significantly higher and lower respectively using the Wilcoxon signed rank test with p<0.05.

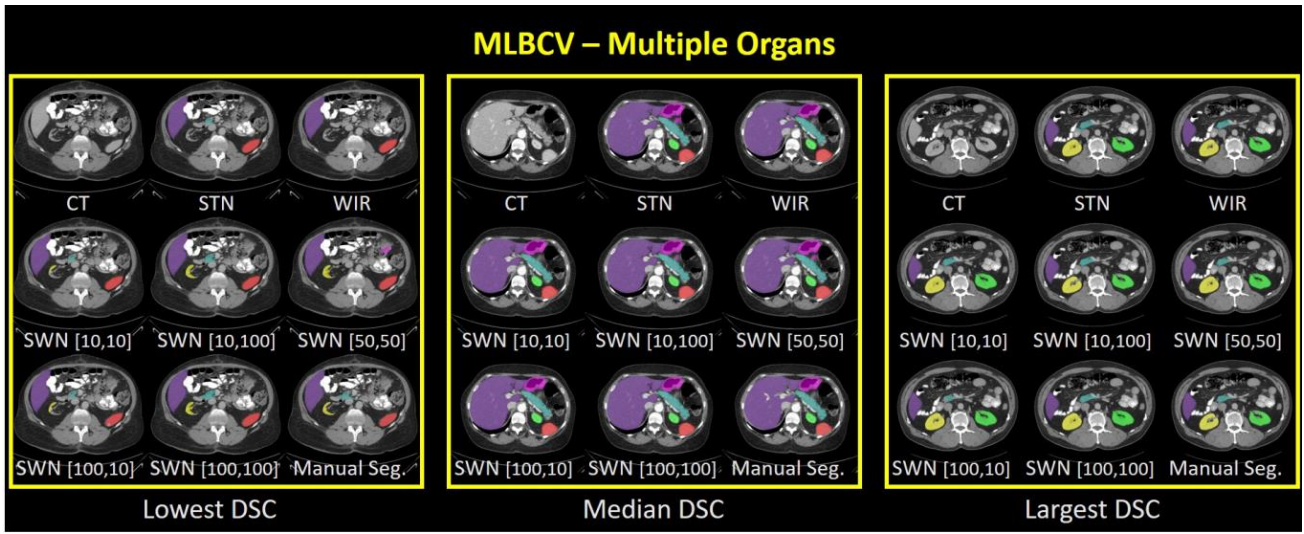

Fig. 6. The qualitative results of applying different intensity normalization strategies. The segmentation results of three scans with the lowest, median, and highest DSC (in SWN [50,50]) are presented for each experiment.

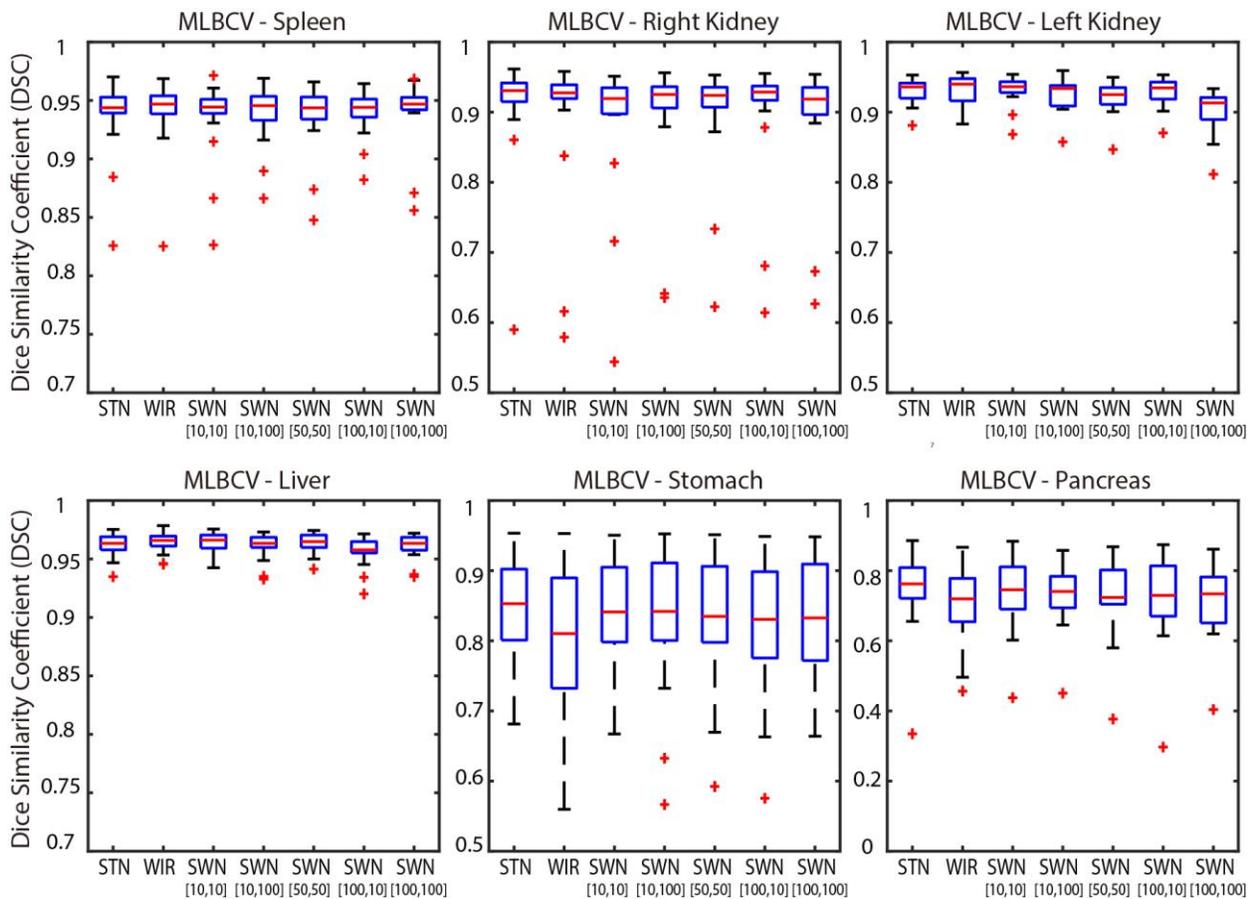

Fig. 7. The quantitative results of applying different intensity normalization strategies to MLBCV dataset, which is from the "same scanner and same population" as training.

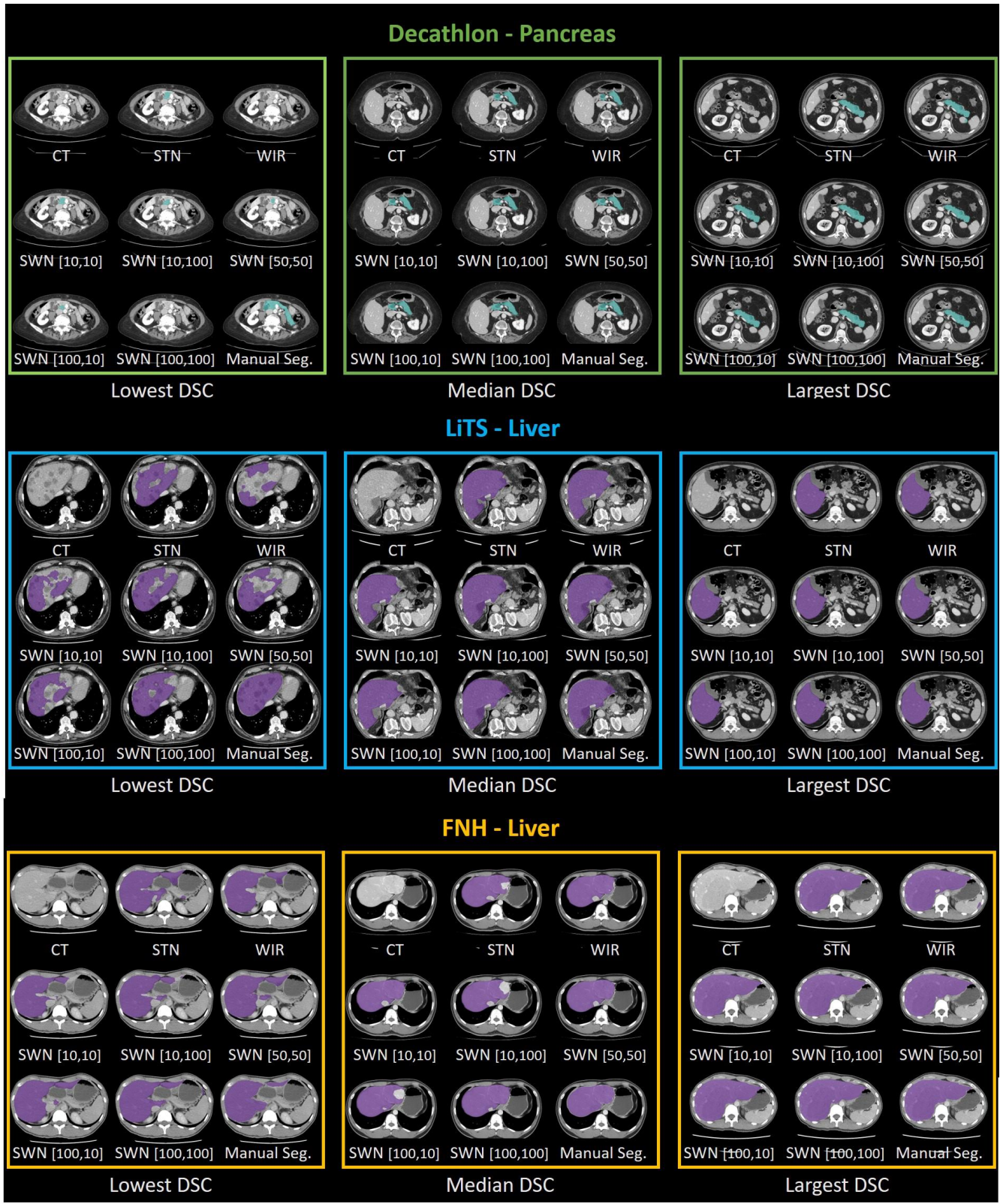

Fig. 8. The qualitative results of applying different intensity normalization strategies. The segmentation results of three scans with the lowest, median, and highest DSC (in SWN [50,50]) are presented for each experiment.

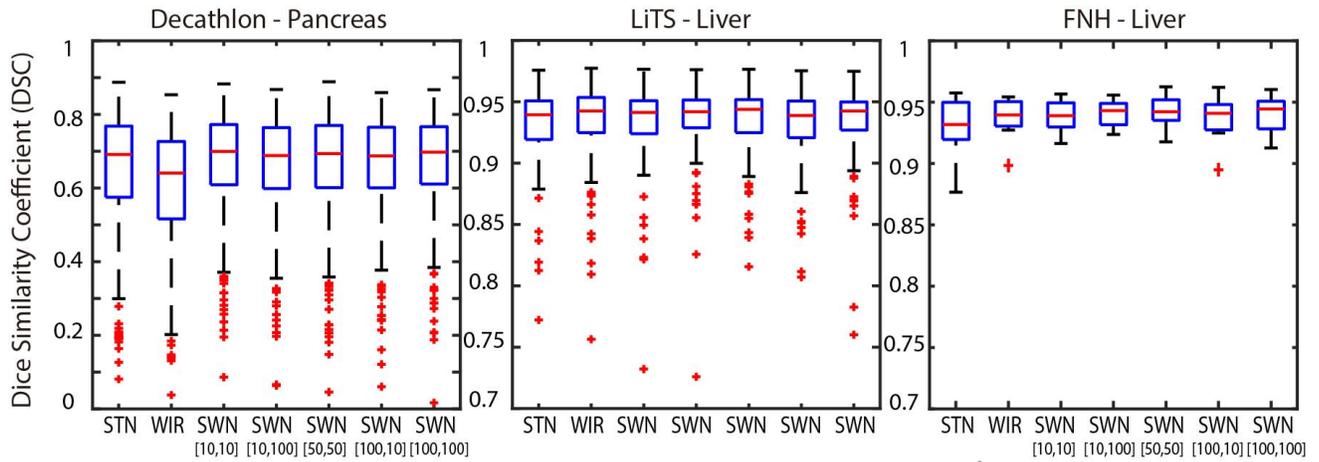

Fig. 9. The quantitative results of applying different intensity normalization strategies to Decathlon, LiTS, and FNH, which are from "same CT contrast, different pathology".

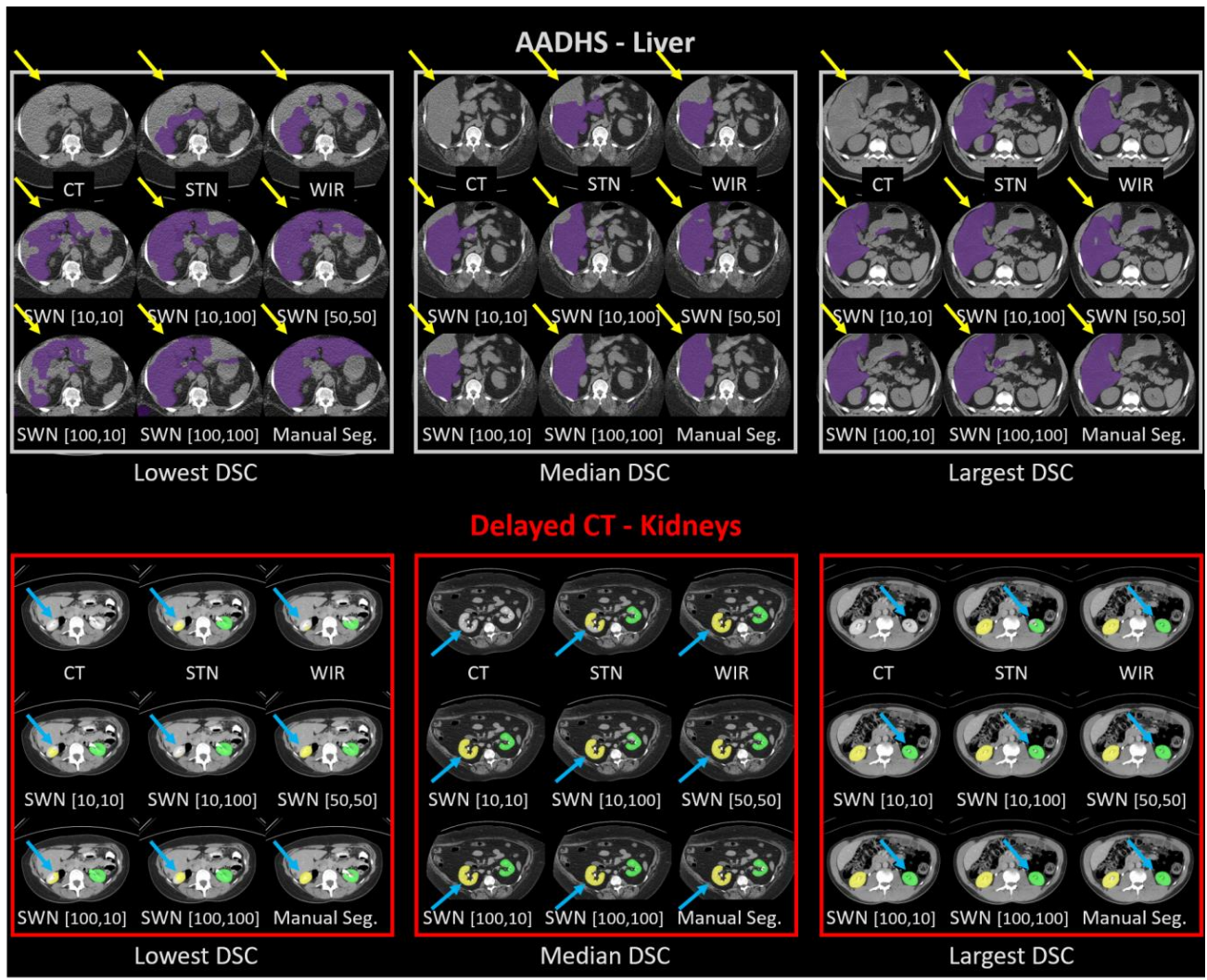

Fig. 10. The qualitative results of applying different intensity normalization strategies on AADHS and Delayed datasets are provided. The segmentation results of three scans with the lowest, median, and highest DSC (in SWN [50,50]) are presented. The yellow and blue arrows indicate the key observations among different methods.

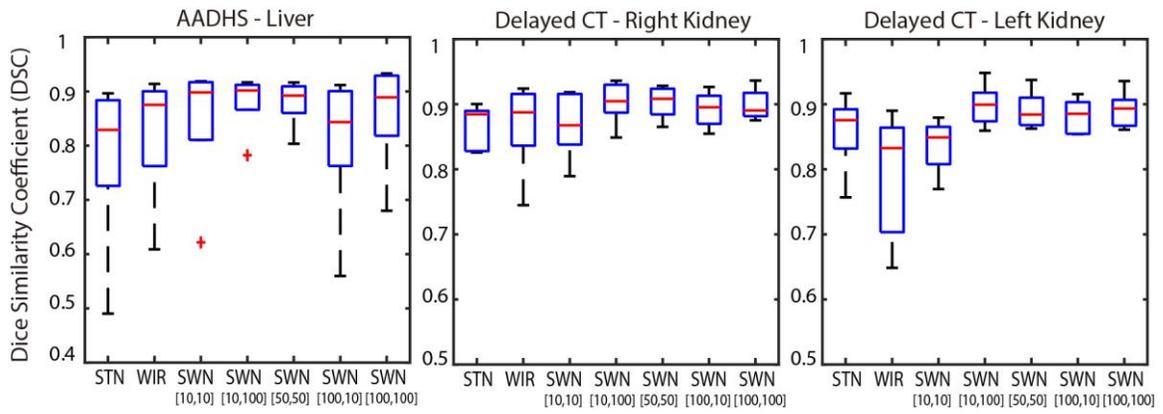

**Fig. 11.** The quantitative results of applying different intensity normalization strategies on the testing scans, which are from "different CT contrast and pathology" compared with training.